\documentclass[twocolumn, trackchanges, twocolappendix]{aastex63}

\usepackage{graphicx}
\usepackage{apjfonts}
\usepackage{amsmath}
\usepackage[raggedright]{subfigure}
\usepackage{color}
\usepackage{hyperref}
\usepackage{acronym}

\def\mesa{\textit{MESA}}

\shorttitle{Magnetic inclination angle evolution of accreting NS in I/LMXB} 
\shortauthors{\sc Yang et al.}

\begin{document}

\title{Magnetic Inclination Evolution of Accreting Neutron Stars in Intermediate/Low-Mass X-ray Binaries}

\author[0000-0001-5532-4465]{Hao-ran Yang}
\author[0000-0002-0584-8145]{Xiang-dong Li}
\affiliation{School of Astronomy and Space Science, Nanjing University, Nanjing 210046, China; lixd@nju.edu.cn}
\affiliation{Key Laboratory of Modern Astronomy and Astrophysics (Nanjing University), Ministry of Education, Nanjing 210023, China}


\begin{abstract} 

\noindent
The magnetic inclination angle $\chi$, namely the angle between the spin and magnetic axes of a neutron star (NS), plays a vital role in its observational characteristics. However, there are few systematic investigations on its long-term evolution, especially for accreting NSs in binary systems. Applying the model of \citet{2021MNRAS.505.1775B} and the binary evolution code \mesa{}, we simultaneously simulate the evolution of the accretion rate, spin period, magnetic field, and magnetic inclination angle of accreting NSs in intermediate/low X-ray binaries (I/LMXBs). We show that the evolution of $\chi$ depends not only on the initial parameters of the binary systems, but also on the mass transfer history and the efficiency of  pulsar loss. Based on the calculated results we present the characteristic distribution of $\chi$ for various types of systems including ultracompact X-ray binaries, binary millisecond pulsars, and ultraluminous X-ray sources, and discuss their possible observational implications.
\end{abstract}


\section{Introduction}

Neutron stars (NSs) in X-ray binaries accrete both mass and angular momentum from their companion stars. If the NSs are magnetized, the interaction between the magnetic field and the accreting material determines the structure of the magnetosphere and the radiation characteristics. The nature of the interaction depends on whether the NSs are wind-fed or disk-fed, usually corresponding to NSs in high-mass X-ray binaries (HMXBs) and intermediate/low X-ray binaries (I/LMXBs), respectively \citep{1991PhR...203....1B}. It is interesting to notice that the torque exerted by the accreting material can simultaneously affect the evolution of the spin period $P_{\rm s}$ of the NSs \citep{1979ApJ...234..296G,1987A&A...183..257W}, the spin inclination angle $\alpha$, and the magnetic inclination angle $\chi$ \citep{1981A&A...102...97W,1990MNRAS.242..188L,2003A&A...404.1023B,2010A&A...520A..76A}. Here $\alpha$ and $\chi$ are the angle between the spin axis of the NS and the axis of the orbital plane and the angle between the spin and magnetic axes of the NS, respectively. In the following we adopt a dipolar configuration for the NS magnetic field.

The magnetic inclination angle plays a crucial role in the characteristics of the pulsed radiation from accreting  X-ray pulsars and from non-accreting  radio pulsars in binaries. The latter are generally related to binary millisecond pulsars (BMSPs), which are thought to evolve from I/LMXBs. Although there are attempts trying to compare theory with observation, definite conclusions are not ready to draw, because reliable data of $\chi$ are still limited \citep{1988MNRAS.234..477L,1990ApJ...352..247R,2010MNRAS.402.1317Y,2009ApJ...707..800V,2014ApJS..213....6J,2021A&A...647A.101B}. Meanwhile, the evolution of the magnetic inclination angles is dependent on the evolution of the accretion rate and the magnetic field of the NS, and the magnetic field-disk interaction, which are currently not well understood.

\citet[hereafter BA21]{2021MNRAS.505.1775B} recently developed a model to trace the NS's magnetic inclination
angle evolution for both disk-fed and wind-fed NSs. They showed that the accretion torque can affect the magnetic inclination angle evolution when both $\alpha$ and $\chi$ significantly deviate from zero. As the spin axis of the NS is being aligned
with the spin-up torque, the magnetic axis becomes misaligned with the spin axis,
which is favorable for detectioin of pulsed radiation from BMSPs.
This work focuses on the magnetic inclination angle evolution for disk-fed NSs in I/LMXBs based on the BA21 model. Here we include some critical factors that were not considered by BA21. The most important one is that BA21 used fixed accretion rates in their calculations for convenience, without considering the influence of binary evolution on the change of accretion rates. So their calculations are limited within $10^5-10^7$ yr evolution. In realistic situation, the accretion history is much more complicated, depending on the initial conditions of both the NSs and the donors, the mass and angular momentum transfer between the components, and the accretion disk physics (see below). The mass transfer lifetimes in I/LMXBs vary from $\sim 10^8$ yr to $10^{10}$ yr, so the final magnetic inclination angles could significantly deviate from those in short-time evolution.

The rest of the paper is organized as follows.
In Section~\ref{s:method}, we review and slightly modify the BA21 model, and then simulate a grid of I/LMXBs with different initial parameters using the binary evolution code \mesa{} \citep{2011ApJS..192....3P,2013ApJS..208....4P,2015ApJS..220...15P,2018ApJS..234...34P,2019ApJS..243...10P}. We select five representative systems, corresponding to  ultracompac X-ray binaries (UCXBs), traditional LMXBs, IMXBs, and ultraluminous X-ray binaries (ULXs), and calculate their evolution. The results are presented in Section~\ref{s:result}.  In Section~\ref{s:discussion}, we discuss the possible effects for the adopted parameters and make predictions for future observational test. Finally, the conclusions are given in Section~\ref{s:conclusion}.

\section{Method}\label{s:method}
\subsection{The evolution of an accreting neutron star}\label{s:model}
 In the BA21 approach, the evolution of an accreting NS can be described by a set of differential equations:
\begin{align}
    I\dot{\Omega} &= n_{1}\cos \alpha + n_{2} + n_{3}(1+\sin^2\chi)-\dot{I}\Omega, \label{eq:omega}\\
    I\Omega\dot{\alpha} &= - n_1\sin\alpha, \label{eq:alpha} \\
    \nonumber I\Omega\dot{\chi} &= \eta\ A(\eta,\alpha,\chi)n_1\sin^2\alpha\cos\alpha \sin\chi \cos\chi \\
    &\quad+n_3\sin\chi \cos\chi, \label{eq:chi}
\end{align}
where $I$ and $\Omega$ are the moment of inertia and angular velocity of the NS, $n_1$, $n_2$ and $n_3$ represent the averaged torques acting on the NS caused by accretion, magnetic braking due to magnetic field-disk interaction, and pulsar's radiation loss, respectively. The coefficient $0<\eta\le 1$ is a constant to describe the accretion torque modulation within the spin period, which is set to unity in accordance with BA21, and $A(\eta,\alpha,\chi)$ is a normalization function,
\begin{equation}
    A(\eta,\alpha,\chi)=\left[1-\frac{\eta}{2}(\sin^2\chi \sin^2\alpha+2\cos^2\chi \cos^2\alpha)\right]^{-1}. \label{eq:A}
\end{equation}
The torques acting on the NS can be written as
\begin{align}
n_{1} &=\dot{M}_{\rm acc}\left(GM_{*}r_{\rm in}\right)^{1/2},\  \mathrm{if} \ r_{\rm in}<r_{\rm co};\label{eq:n1}\\
n_{2} &=-\frac{\mu^2}{3r_{\rm co}^3},\ \mathrm{if} \ r_{\rm in}<r_{\rm lc};\label{eq:n2}\\
n_{3} &= -\frac{\mu^2}{r_{\rm lc}^3} \label{eq:n3}.
\end{align}
Here $G$ is the gravitational constant, $M_{*}$, $\mu$ and $\dot{M}_{\rm acc}$ are the mass, magnetic moment, and mass accretion rate of the NS, respectively. We assume that the accretion rate is the mass transfer rate $\dot{M}$ limited by the Eddington limit accretion rate, that is
\begin{equation}
    \dot{M}_{\rm acc} = \min(\dot{M},\dot{M}_{\rm Edd}), \label{eq:acc1}
\end{equation}
where
\begin{equation}
\dot{M}_{\rm Edd}=1.43\times10^{18}M_{*,\odot} \rm g\,\rm s^{-1}, \label{eq:edd}
\end{equation}
and $M_{*,\odot}=M_{*}/\rm M_{\odot}$.

In Eqs.(\ref{eq:n1})-(\ref{eq:n3}), $r_{\rm in}$, $r_{\rm co}$ and $r_{\rm lc}$ are the inner disk radius (or the  magnetospheric radius), corotation radius, and light cylinder radius, respectively,
\begin{align}
\nonumber r_{\rm in}&=\xi\left(\frac{\mu^4}{2GM_*\dot{M}^2}\right)^{1/7}\\
         &\simeq7.5\times10^7\mu_{30}^{4/7}M_{*,\odot}^{-1/7}(\dot{M}/\dot{M}_{\rm Edd,\odot})^ {-2/7}\rm{cm}\label{eq:rm}\\
r_{\rm co}&=\left(\frac{GM_*}{\Omega^2}\right)^{1/3}\simeq1.5\times10^{8}M_{*,\odot}^{1/3}P_{\rm s,1}^{2/3}\rm{cm},\label{eq:rco}\\
r_{\rm lc}&=\frac{c}{\Omega}\simeq4.8\times10^{9}P_{\rm s,1}\rm{cm}. \label{eq:rlc}
\end{align}
where $\xi\sim 0.5$ is a correction coefficient depending on the detailed structure of the disk \citep{1979ApJ...234..296G,2005ApJ...634.1214L,2008A&A...478..155B}, $\mu_{30,0}=\mu /10^{30}\,\rm G\,cm^3$ and $P_{\rm s,1}=P/1\rm s$ are the normalized magnetic dipole moment and spin period of the NS, respectively.
Equation (10) is derived by assuming that the total matter (ram and gas) pressure is balanced by the total magnetic pressure at $r_{\rm in}$, which is generally consistent with axisymmetric and global 3D MHD simulations \citep{2013MNRAS.433.3048K}. Since there is  no specific restriction on the geometry of the stellar magnetic field in this formula \citep{2015SSRv..191..339R}, it has also been adopted in more complicated situations, including those that are non-stationary and with tilted magnetic and rotational axes \citep[e.g.,][]{2021MNRAS.506..372R}.

We also consider the accretion-induced field decay in the following form \citep{s89,zk06,2019RAA....19...44L},
\begin{equation}
    \mu =\mu_{\rm min}+\mu_{0}\left(1+\frac{\Delta M_*}{10^{-5}\rm{M}_{\odot}}\right)^{-1}.\label{eq:mu}
\end{equation}
where $\mu_0$,  $\mu_{\rm min}$, and $\Delta M_*$ are the initial magnetic moment, the bottom magnetic moment, and the amount of matter accreted by the NS, respectively. We set $\mu_{\rm min}=10^{26}\,\rm G\,cm^3$, to be comparable with the weakest magnetic fields of pulsars.

\subsection{Evolution of I/LMXBs}\label{s:mesa}

\begin{table}
\setlength{\tabcolsep}{4mm}
\caption{Parameters of the selected models in this paper. $M_{\rm 2,i}$ and $P_{\rm orb,i}$ represent the initial donor mass and orbital period, respectively. The initial mass of NS is set to $1.4\,\rm M_{\odot}$ for all models.}
\label{tab:mesa}
\centering
\begin{tabular}{lccccc}
\hline
Model & A & B & C & D & E \\  
\hline                        
$M_{\rm 2,i}\,(\rm M_{\odot})$ & 1 & 1 & 3.4 & 3 & 5 \\
$P_{\rm orb,i}\,(\rm d)$ & 1 & 10 & 1.58 & 100 & 1 \\
\hline                                   
\end{tabular}
\end{table}

\begin{figure*}[t]
	\centering
	\includegraphics[width=\textwidth]{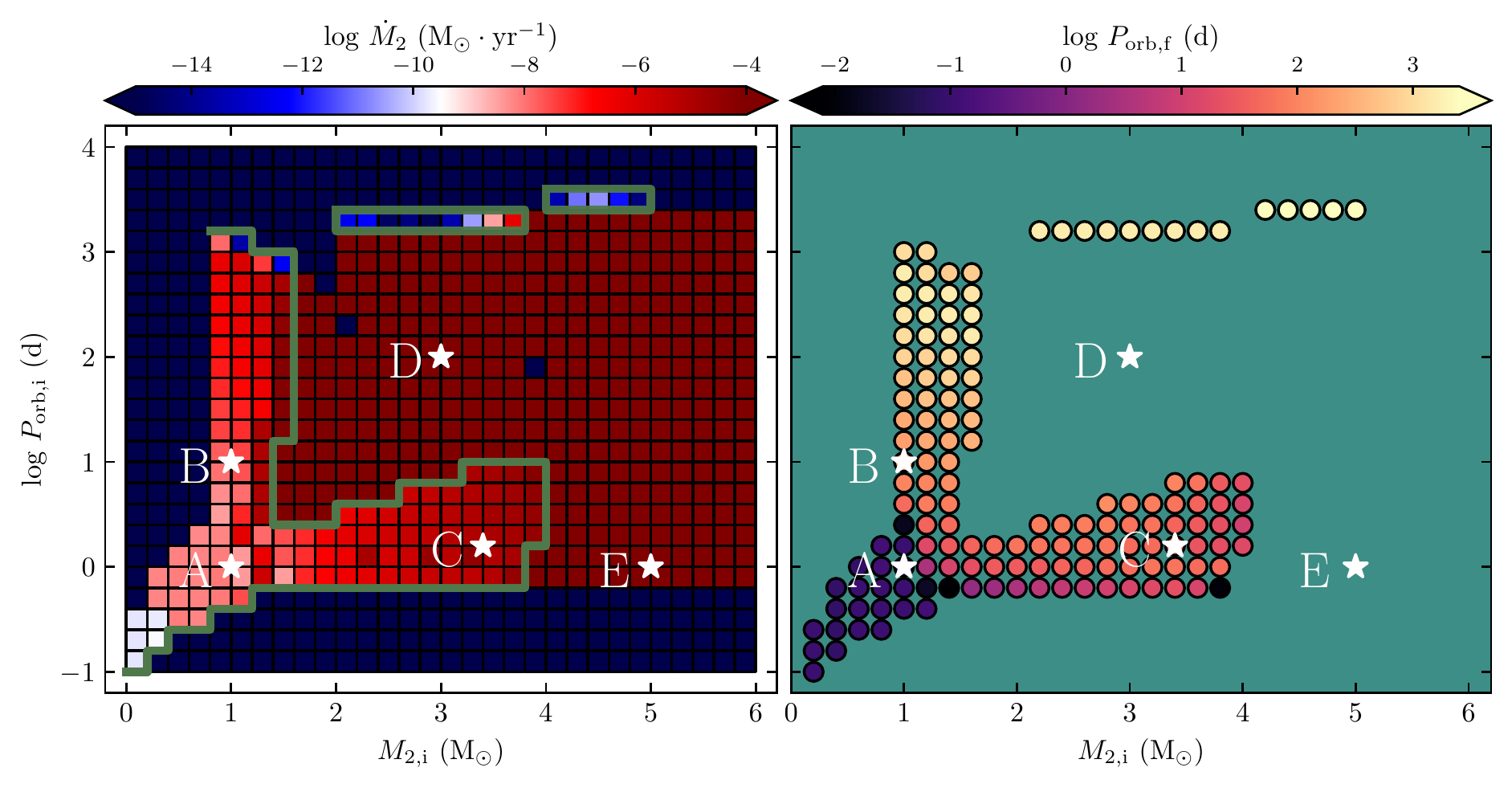}
	\caption{Distributions of I/LMXBs in the $M_{\rm 2,i}-P_{\rm orb,i}$ plane, colored by the magnitude of the maximum mass transfer rate $\dot{M}_2$ (left panel) and the final orbital period $P_{\rm orb,f}$ (right panel). The green lines in the left panel confine the parameter space that have stable mass transfer. The white stars in both panels are the selected systems used in our calculations, and their parameters are listed in Table~\ref{tab:mesa}.}
	\label{fig:mesa}
\end{figure*}

\begin{figure*}[!t]
	\centering
	\includegraphics[width=\textwidth]{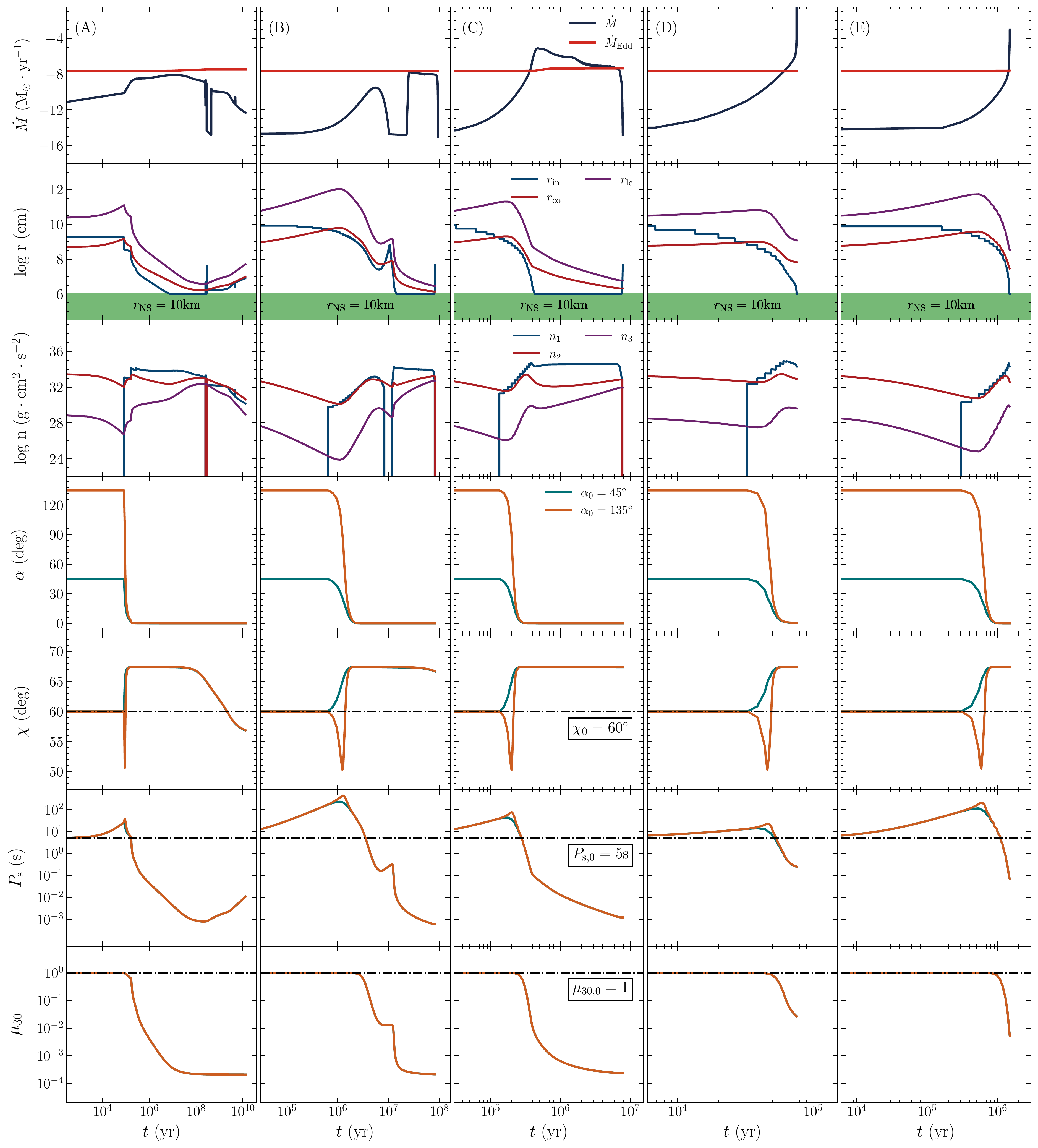}
	\caption{Evolution of different parameters for  systems A-E (from left to right). The rows from top to bottom show the change of the mass transfer rates, the three radii ($r_{\rm in}$, $r_{\rm co}$ and $r_{\rm lc}$), three torques, the spin inclination angle $\alpha$, the magnetic inclination angle $\chi$, the spin period $P_{\rm s}$, and the magnetic moment $\mu_{30}$ with time.}
	\label{fig:reference}
\end{figure*}

We follow the long-term binary evolution using the stellar evolution code \mesa{} (version number 15140). The NS is taken to be a point mass whose initial mass is set to $1.4\,\rm M_{\odot}$, and the metallicity of the donor is $Z=0.02$. We evolved a large number of incipient NS I/LMXBs with the donor mass varying from $0.2\,\rm M_{\odot}$ to $6\,\rm M_{\odot}$ by steps of $0.2\,\rm M_{\odot}$, and the orbital period (in units of days) logarithmically ranging from $-1$ to 3 by steps of 0.2. The \citet{1988A&A...202...93R} scheme was used to compute the mass transfer rates via Roche-lobe overflow (RLOF). Besides, we considered magnetic braking and gravitational wave radiation for angular momentum loss from the binary.

Figure~\ref{fig:mesa} illustrates the distribution of I/LMXBs in the initial donor mass-orbital period plane. In the left panel, the color of each element represents the maximum mass transfer rate $\dot{M}_2$, the magnitude of which is indicated by the color bar on the top of the figure. If the maximum mass transfer rate is larger than $10^{-4}\,\rm M_{\odot}\, \rm yr^{-1}$, we regard the mass transfer to be dynamically unstable, followed by common envelope (CE) evolution; if the maximum mass transfer rate is smaller than $10^{-15}\,\rm M_{\odot}\,\rm yr^{-1}$ we regard that no mass transfer via RLOF has occurred. Systems in between are confined by the green lines, and identified to experience stable mass transfer. They are plotted in the right panel with the color denoting the magnitude of their final orbital periods $P_{\rm orb,f}$, also indicated by the top color bar.
The white stars labeled A-E in both panels are representative systems used in our following calculations whose initial parameters are listed in Table~\ref{tab:mesa}. The five systems have distinct evolutionary paths, three of which (A, B and C) are located in the parameter space of stable mass transfer and the other two (i.e. D and E) are outside the parameter space.

Systems A and B are  LMXBs with the same initial donor mass ($1\,\rm M_{\odot}$) but different initial orbital periods. They will follow different evolutionary paths. In system A, because of its relatively short orbital period (1 d), mass transfer is driven by orbital angular momentum loss caused by magnetic braking and gravitational wave radiation. It starts early and lasts $\sim 10^{10}$ yr, and the donor always remains in main-sequence. The binary will finally evolve to to be an UCXB (and probably a black widow/redback binary). The initial orbital period of system B is 10 d, longer than the bifurcation period which separates the converging binary systems from the diverging binary systems \citep{1988A&A...191...57P,1989A&A...208...52P}. Its mass transfer is driven by nuclear evolution of the donor, which will finally evolve to be a Helium white dwarf (He WD). The duration ($\sim 10^{8}$ yr) of the mass transfer is significantly shorter than in System A. After the mass transfer ceases, the final evolutionary outcome is a BMSP.

Systems C, D and E start evolution as IMXBs and can appear as ULXs. Among them, system C experiences thermal-timescale mass transfer at first, and then evolves to be an LMXB after the mass ratio reverses. The donor eventually becomes a hybrid Carbon-Oxygen white dwarf (CO WD). The remaining two systems (D and E) are subject to delayed dynamically unstable mass transfer because the initial orbital period is too long (100 d) or the donor is too massive ($5\,\rm M_{\odot}$). The mass transfer rates rise rapidly to exceed the Eddington limit accretion rate, and they will enter CE evolution, probably leading to merger of both components.

Knowing the mass transfer history, we can follow the evolution of the spin period and the magnetic inclination angle for given initial parameters. It is potentially possible to predict the
distribution of the magnetic inclination angles of the NSs in different evolutionary stages.

\section{Results}\label{s:result}
\begin{figure*}[!t]
	\centering
	\includegraphics[width=\textwidth]{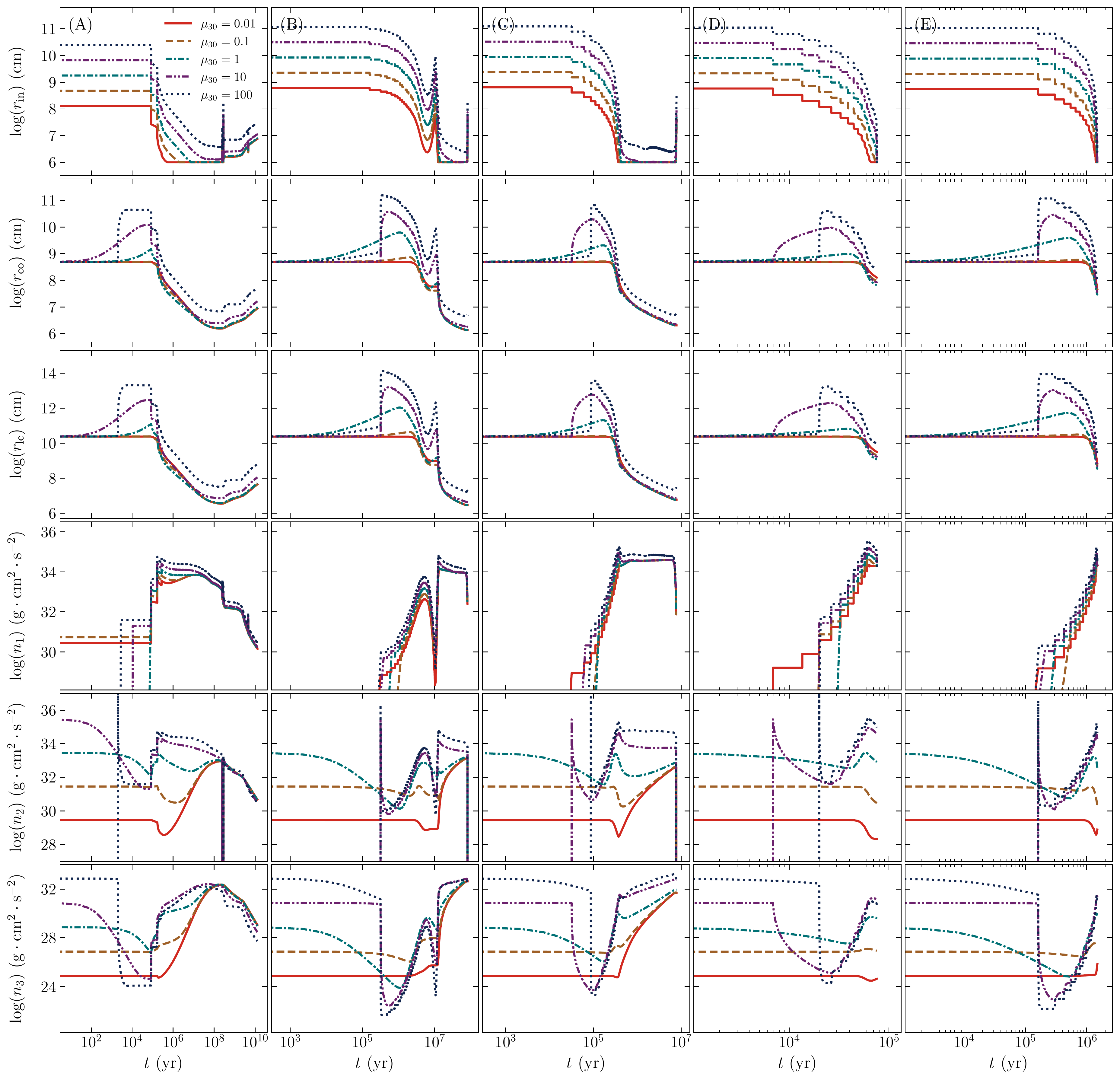}
	\caption{Evolution of the three radii and the three torques with different initial magnetic moments. The other initial parameters are $\alpha_0=45^{\circ}$, $\chi_0=60^{\circ}$, and $P_0=5\rm s$.}
	\label{fig:magn}
\end{figure*}

\begin{figure*}[!t]
	\centering
	\includegraphics[width=\textwidth]{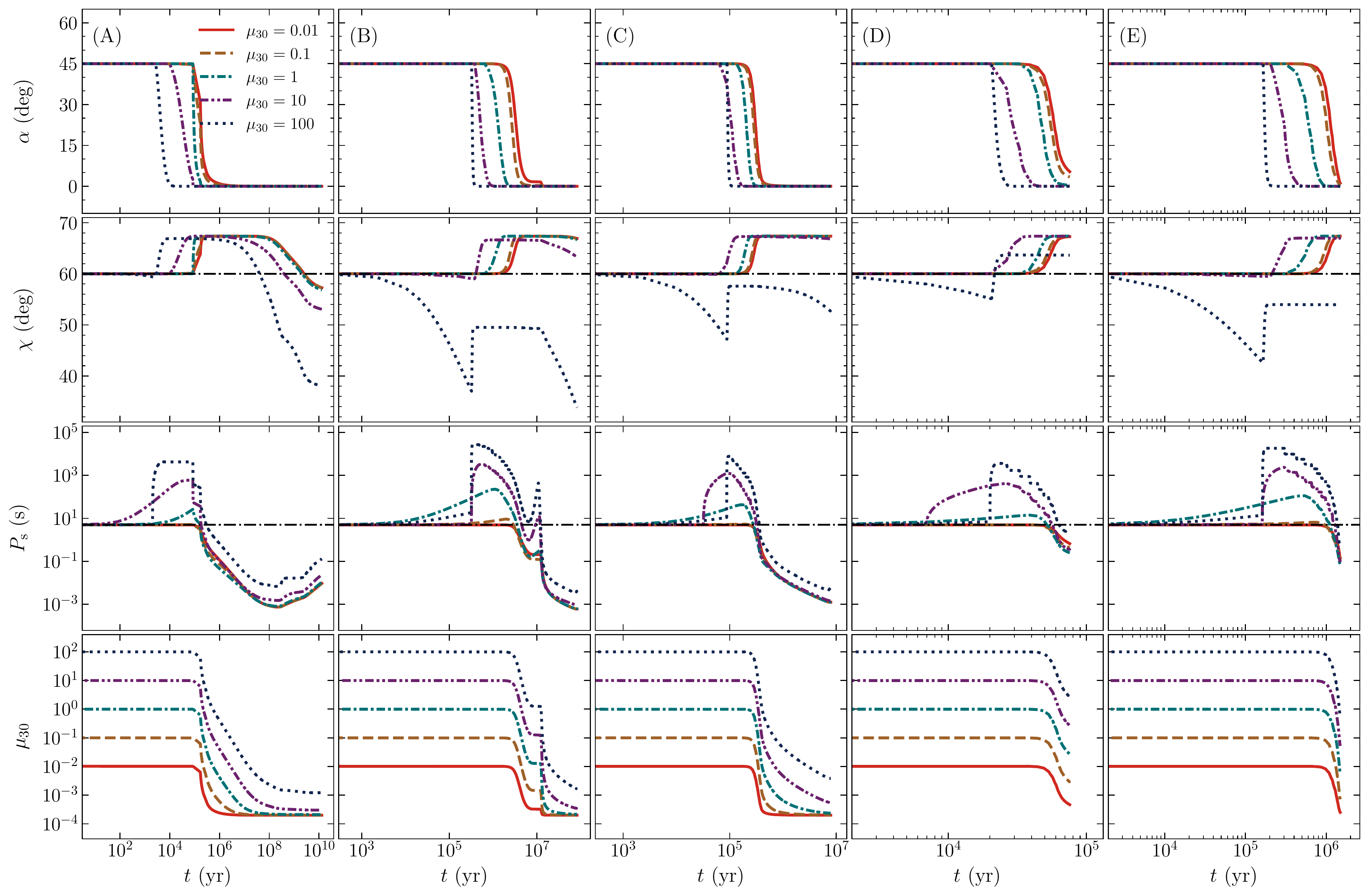}
	\caption{Same as Fig.~\ref{fig:magn}, but for the evolution of the spin inclination angle $\alpha$, the magnetic inclination angle $\chi$, the spin period $P_{\rm s}$, and the magnetic moment $\mu_{30}$.}
	\label{fig:mag}
\end{figure*}

\begin{figure*}[!t]
	\centering
	\includegraphics[width=\textwidth]{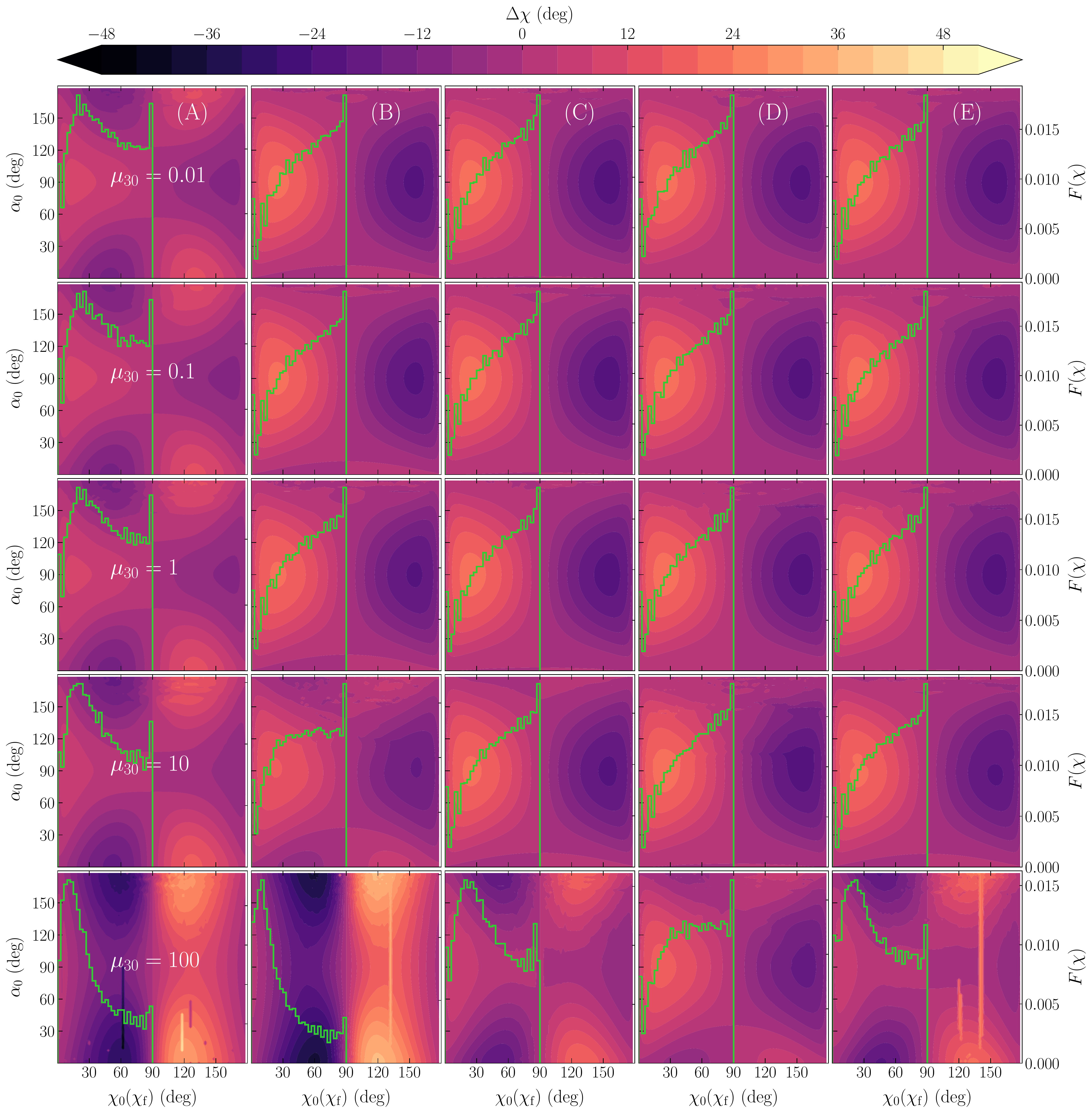}
	\caption{The variations of the magnetic inclination angles $\Delta\chi$ for the five systems with different initial magnetic moments in the $\chi_0-\alpha_0$ plane. The green lines represent the probability distribution of the final magnetic inclination angles $\chi_{\rm f}$.}
	\label{fig:chif}
\end{figure*}

Figure~\ref{fig:reference} shows the calculated results (in solid lines) for the five systems. The initial parameters at the beginning of mass transfer are taken to be in accordance with BA21 for the convenience of comparison: the magnetic inclination angle $\chi_0=60^{\circ}$, the spin inclination angle $\alpha_0=45^{\circ}$ and $135^{\circ}$ (indicated by the blue and orange lines, respectively), the spin period\footnote{Before the mass transfer occurs, the NS is slowed down to a relatively long spin period due to magnetic dipole radiation. We have performed calculations with $P_{\rm s,0}$ ranging from 10 ms to 10 s, and found that the results are insensitive to the value of  $P_{\rm s,0}$.} $P_{\rm s,0}=5\ \rm s$, and the normalized magnetic dipole moment $\mu_{30,0}=\mu_0 /10^{30}\,\rm G\,cm^3=1$. The panels from top to bottom in Figure~\ref{fig:reference} demonstrate the evolution of the mass transfer rate, the three radii ($r_{\rm in}$, $r_{\rm co}$ and $r_{\rm lc}$), the three torques ($n_1$, $n_2$ and $n_3$), the spin inclination angle $\alpha$, the magnetic inclination angle $\chi$, the spin period $P_{\rm s}$, and the magnetic moment $\mu_{30}$, respectively. The five rows from left to right correspond to systems A to E, respectively.

We can easily find that the evolution of $\chi$ with different $\alpha_0$ shows similar tendency except in the beginning phase of the mass transfer, which is related to the first term in Eq~(\ref{eq:chi}) ($\propto \sin^2\alpha\cos\alpha$). Moreover, the evolution of $\chi$ is similar for all systems except system A. During the first $10^4 - 10^5\,\rm yr$ of mass transfer, the mass transfer rates $\dot{M}$ are relatively low and there is no or very little accretion, thus little change occurs in both $\alpha$ and $\chi$. The NS is spun down mainly by the magnetic braking torque $n_2$. Then, with the decrease in $\Omega$ and increase in $\dot{M}$, $n_2$ declines and $n_1$ increases accordingly. Therefore, $\alpha$ starts evolving toward $0\degr$ rapidly (within a few $10^4-10^6$ yr). During this period, $\chi$ increases to $\sim 68\degr$ (in the case of $\alpha_0=135\degr$, $\chi$ decreases firstly because $\cos\alpha<0$), and keeps nearly unchanged after $\alpha\rightarrow 0\degr$. At the same time, the accreting NS is spun up by the accretion torque $n_1$. Our calculated evolution of $\chi$ in these circumstances is in general concordance with the results in BA21. However, we note that the long-term spin evolution of the NS in systems B-E is different despite of the similarity of the $\chi$ evolution. The spin periods $P_{\rm s}$ in system B and C evolve to milliseconds eventually after the accretion ends due to their relatively long accretion time and low magnetic moments ($\sim \mu_{\rm min}$). The latter reflects effective accretion-induced field decay occurred in the NS. For the other two systems (D and E), the shorter mass transfer time leads to less decayed magnetic moment and a longer $P_{\rm s}$. The mass transfer rates in the late stage exceed the Eddington limit by a few orders of magnitude, so these systems would behave as ULXs.

As for system A, the accretion timescale is so long (up to the Hubble time) that the evolution of $\chi$ enters another stage. From Eq.~(\ref{eq:chi}) we know that the evolution of $\chi$ depends on the accretion torques $n_1$ and the pulsar loss torque $n_3$. After $\alpha$ evolves to zero, the change in $\chi$ is completely controlled by $n_3$. Although its magnitude is relatively small, as the mass transfer proceeds for sufficiently long time, $n_3$ drives $\chi$ to decrease to $\sim 55\degr$.

To explore the influence of the magnetic moment, we set the initial parameters to be $\alpha_0=45^{\circ}$, $\chi_0=60^{\circ}$ and $P_0=5\,\rm s$, and change $\mu_{30,0}$ from 0.01 to 100. This range roughly covers the magnetic field strengths of weakly magnetized NSs to magnetars. Figure~\ref{fig:magn} shows the evolution of the three radii ($r_{\rm in}$, $r_{\rm co}$ and $r_{\rm lc}$) and the magnitude of the three torques ($n_1$, $n_2$ and $n_3$) with different $\mu_{30,0}$, and Figure~\ref{fig:mag} shows the corresponding evolution of $\chi$, $\alpha$, $P_{\rm s}$ and $\mu_{30}$.

 We first examine the evolution of system A, which is demonstrated in the first column in Figures~\ref{fig:magn} and~\ref{fig:mag}. When $\mu_{30,0}\ge 1$, $r_{\rm in}$ is comparable with $r_{\rm co}$ at the beginning of the mass transfer, and the NS experiences an episode of spin-down. The longest period that the NS can reach increases with $\mu_{30,0}$. After that the NS enters the accretor phase and the accretion torque $n_1$ begins to work. We note that the stronger $\mu_{30,0}$ is, the earlier the accretor phase starts, and $\alpha$ and $\chi$ begin to decrease and increase earlier, respectively. After about $10^6\,\rm yr$, the mass transfer rate gets higher, accretion causes $\mu_{30}$ to decrease significantly, leading to smaller $r_{\rm in}$ and stronger $n_1$. The NS starts spinning up rapidly.  Accordingly, both $r_{\rm co}$ and $r_{\rm lc}$ become smaller, resulting in more effective spin-down torques ($n_2$ and $n_3$) to balance $n_1$, which not only set the spin period to be around the equilibrium spin period\footnote{After the NS reaches the equilibrium period, $r_{\rm in}$ alternates between $\lesssim r_{\rm co}$ and $\gtrsim r_{\rm co}$, and the accretion torque $n_1$ correspondingly switches between zero and non-zero values according to Eq.~(5). In Figure~3 we only show the non-zero values of $n_1$ for clarity.}, but also cause $\chi$ to decrease. Because the magnitude of $n_3$ increases with $\mu_{30,0}$, the larger $\mu_{30,0}$, the smaller the final $\chi$. On the other hand, if the initial magnetic field is relatively weak ($\mu_{30,0}<1$), the NS directly enters the accretor phase at the beginning of the mass transfer. Smaller magnetic moments lead to less change in $\chi$. Similar tendency can be found in systems B-E,
except the case of $\mu_{30,0}=100$, in which the initial inner disk radius is beyond the light cylinder radius, and the NS enters the radio pulsar phase at the beginning. This causes the magnetic inclination angle $\chi$ to decrease for $\sim 10^4 - 10^5\,\rm yr$ before accetion begins. After that, $\chi$ increases and then settles down. If the mass transfer lasts sufficiently long time, $\chi$ will finally decreases. The early radio pulsar episode leads to the smallest $\chi$ after accretion ends in the five systems.

We then calculate the evolution for the five systems with  $\alpha_0$ and $\chi_0$ randomly distributed between $0^{\circ}$ and $180^{\circ}$. Figure~\ref{fig:chif} shows the variations of the magnetic inclination angles $\Delta\chi$ in the $\chi_0-\alpha_0$ plane. Each panel corresponds to a system with a designated initial magnetic dipole moment $\mu_{30,0}$. We also depict the probability distribution of the final magnetic inclination angles $\chi_{\rm f}$ with the green histogram in each panel (to be discussed below).  It is apparent to see the upper and lower symmetry and mirror symmetry in left and right due to the spherical symmetry of the NS. And as mentioned above, in most cases the magnetic axis tends to misalign with the spin axis of the NSs except for system A and for some cases in other systems with $\mu_{30,0}\ge 100$, in which the magnetic axis tends to align with the spin axis.

\section{Discussion}\label{s:discussion}
\subsection{The effect of transient accretion}\label{s:di}
\begin{figure*}[!t]
	\centering
	\includegraphics[width=\textwidth]{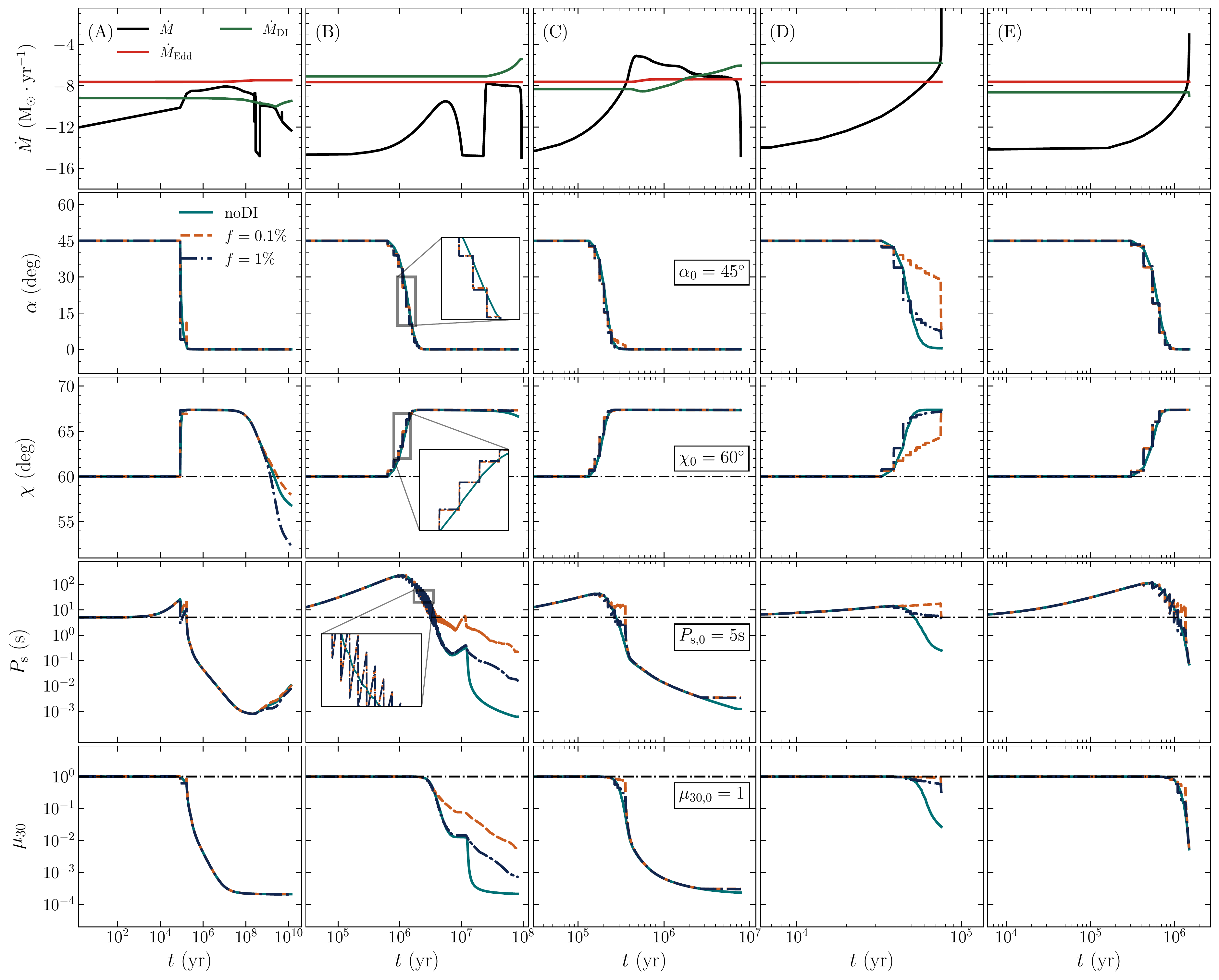}
	\caption{The solid and dashed lines compare the results in the reference model and that with the disk instability considered. The initial parameters are $\alpha_0=45^{\circ}$, $\chi_0=60^{\circ}$, $P_0=5 \rm s$, and $\mu_{30,0}=1$. }
	\label{fig:di}
\end{figure*}

In last section we infer the NS accretion rate from the mass transfer rate in I/LMXBs to trace the evolution of the NS. However, it is well known that most LMXBs are transients with rapid accretion during short outbursts separated by long quiescence. The origin of the transient behavior is likely related to the thermal and viscous instability, which occurs when the mass transfer rate $\dot{M}$ is below a critical value \citep{2001NewAR..45..449L},
\begin{align}
    \nonumber\dot{M}_{\rm cr}&\simeq3.2\times10^{-9}\left(\frac{M_{*}}{1.4\rm{M}_{\odot}}\right)^{0.5}\left(\frac{M_{\rm d}}{1.0\rm{M}_{\odot}}\right)^{-0.2}\\
    &\times\left(\frac{P_{\rm orb}}{1.0\rm d}\right)^{1.4}\rm M_{\odot}\, yr^{-1}. \label{eq:cr}
\end{align}
Limit cycles of the accretion rate in the disk results in the transition from quiescence to outburst when the disk gets hot enough and hydrogen is ionized from a cooler and predominantly hydrogen neutral state.
This means that a transient NS would attain a higher accretion rate during outbursts than the long-term average mass transfer rate. With that in mind, the accretion rates should be reformulated as:
\begin{align}
\dot{M}_{\rm acc} =
\begin{cases}
\dot{M}_{\rm di},  & \mathrm{if} \ \dot{M}\le\dot{M}_{\rm cr} \\
\min(\dot{M},\dot{M}_{\rm Edd}),  & \mathrm{if} \ \dot{M}>\dot{M}_{\rm cr}, \\
\end{cases}\label{eq:acc2}
\end{align}
where $\dot{M}_{\rm di}$ is the accretion rate when the disk is subject to the thermal and viscous instability. We simply assume that the accretion rate is enhanced to $1/f$ times of $\dot{M}$ during outbursts for a given outburst duty cycle $f$, and declines to zero at quiescence \citep[see also][]{2017ApJ...835....4B}, that is
\begin{align}
\dot{M}_{\rm di} =
\begin{cases}
\dot{M}_{\rm burst}=\min\left(\frac{\dot{M}}{f},\dot{M}_{\rm Edd}\right),  & \mathrm{during\ outbursts} \\
\dot{M}_{\rm qu}=0,  & \mathrm{during\ quiescence}. \\
\end{cases}\label{eq:di}
\end{align}

We recalculate the evolution of the five systems with disk instability considered. The initial parameters are set to be: $\alpha_0=45^{\circ}$, $\chi_0=60^{\circ}$, $P_0=5\, \rm s$, and $\mu_{30,0}=1$. The results are compared with the reference model with the same parameters but without disk instability considered (in solid line) in Figure~\ref{fig:di}. The dashed and dot-dashed lines correspond to $f=0.1\%$ and $1\%$, respectively. And the insets in the second row demonstrate the detailed evolution of $\alpha$, $\chi$ and $P_{\rm s}$ when disk instability is considered.

Except for system B, which is subject to the disk instability during the whole mass transfer process, all systems generally experience the transient behavior during the early evolutionary stage when $\dot{M}$ is low and rising (and during the late stage for system A when $\dot{M}$ is declining).
It is interesting to compare the magnetic inclination evolution when the mass transfer rate $\dot{M}<\dot{M}_{\rm cr}$. In the case that disk instability is not considered,  the accretion torque $n_1$ is relatively small or even  zero (if $r_{\rm in}>r_{\rm co}$), so there would be little or no change in $\alpha$, and $\chi$ decreases mainly due to magnetic dipole radiation according to Eqs.~(2) and (3). If the disk instability is taken into account, the enhancement of the accretion rate during outbursts would exert an efficient torque on the NS, causing both $\alpha$ and $\chi$ to evolve more rapidly and earlier. We can estimate the time-averaged accretion torque $\left\langle n_1\right\rangle$ to be
\begin{align}
    \nonumber\left\langle n_1\right\rangle &= \dot{M}_{\rm burst}(GM_{*}r_{\rm in, burst})^{1/2}\cdot f\\
    &= n_1(\dot{M})\left(\frac{r_{\rm in, burst}}{r_{\rm in}}\right)^{1/2}
\end{align}
where $r_{\rm in, burst}$ is the inner disk radius during outbursts.

For systems B-E, considering the disk instability does influence the evolutionary paths of $\chi$, but barely affects the magnitude of $\chi_{\rm f}$. However, the disk instability has an important impact on the evolution of the spin period and the magnetic moment, especially for wide binaries such as system B, where the mass transfer rates are smaller than the criteria all the time and then $\left\langle n_1\right\rangle$ is always $< n_1$. Consequently less material is accreted by the NS and less field decay occurs.

\subsection{The effect of pulsar loss torque enhancement}\label{s:enhancement}
\begin{figure*}[!t]
	\centering
	\includegraphics[width=\textwidth]{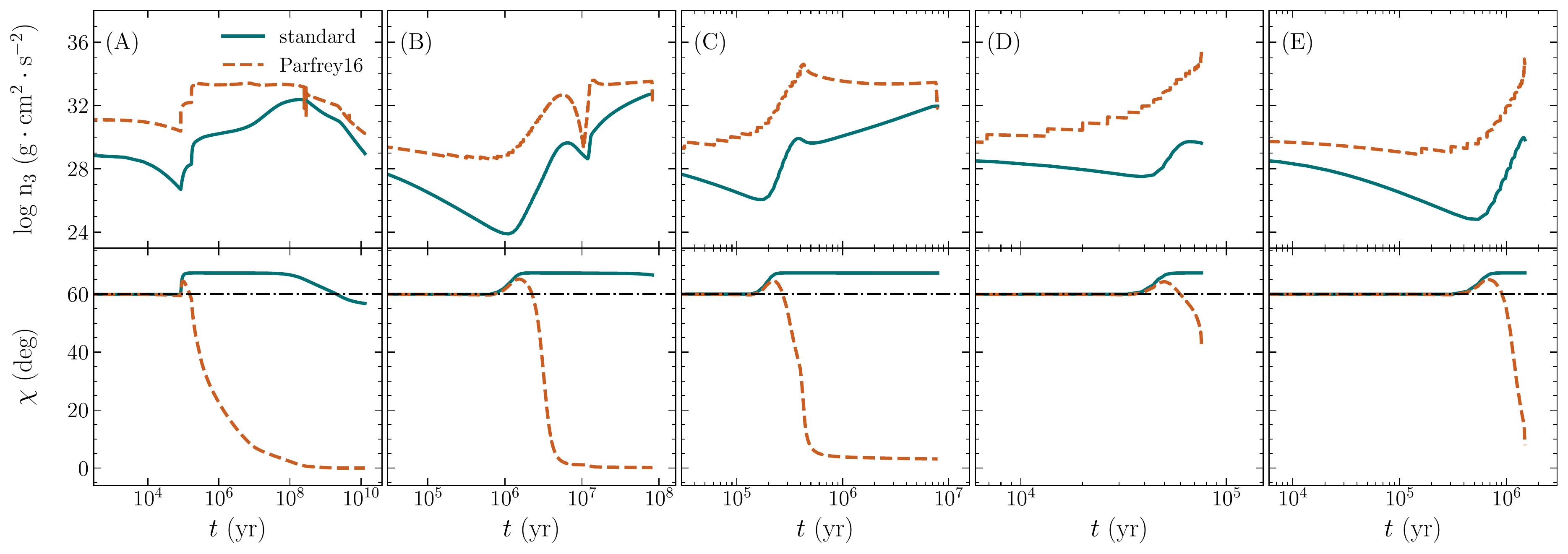}
	\caption{The solid and dashed lines compare the evolution of the magnetic inclination angles without and with torque enhancement considered. The initial parameters are same as in Figure~\ref{fig:di}. }
	\label{fig:pafrey16}
\end{figure*}

\citet{2016ApJ...822...33P} proposed that the presence of a conducting disk around the NS can increase the number of open magnetic field lines and pulsar loss, in particular for rapid rotators. Under this circumstance, the torque $n_3$ related to pulsar loss is amplified to be
\begin{align}
    \nonumber n_{\rm3, parf} &=\left(\zeta\frac{r_{\rm lc}}{r_{\rm in}}\right)^2n_3\\
    &\simeq -1.65\times10^{33}\dot{M}_2^{4/7}M_{*,\odot}^{2/7}\mu_{30}^{6/7}\Omega\ \rm{g\, cm^2\, s^{-2}},
\end{align}
where $\zeta$ is used to describe the efficiency of the field line opening by the differential rotation between NS and the disk, and we take $\zeta=1$. Therefore, the total toque exerted on the NS should be rewritten as:
\begin{align}
n_{\rm tot} =
\begin{cases}
n_1+n_2+n_{\rm parf},  & \mathrm{if} \ r_{\rm in}<r_{\rm co} \\
n_2+n_{\rm parf},  & \mathrm{if} \ r_{\rm co}\le r_{\rm in}<r_{\rm lc} \\
n_3, & \mathrm{if} \ r_{\rm in}\ge r_{\rm lc}.
\end{cases}\label{eq:parf}
\end{align}
We recalculate the evolution with the initial parameters same as in the previous section and present the results in Figure~\ref{fig:pafrey16}. The upper panel compares the torque $n_3$ and the lower panel shows its effect on the evolution of the magnetic inclination angles $\chi$ for the five systems. It is evident that the torque enhancement plays an important role in the late evolution of $\chi$. Consequently, the magnetic axis tends to align with the spin axis, which is basically consistent with the conclusion by BA21.

\subsection{Predictions of the $\chi$ distributions}\label{s:observe}

Based on the above calculations, we attempt to predict the possible distributions of the magnetic inclination angles for NSs evolved from I/LMXBs. We recall that the five selected systems (A-E) correspond to different evolutionary outcomes of NS I/LMXBs, namely X-ray pulsars in UCXBs (A), BMSPs with He WD companions (B) and with CO WD companions (C), and ULXs (D and E).

In Figure~\ref{fig:chif}, we depict the probability distribution functions $F_{\chi}$ of the final magnetic inclination angles $\chi_{\rm f}$ denoted by $F_{\chi}=\frac{1}{N}\int n_{\chi} \, d{\chi}$ assuming that $\alpha_0$ and $\chi_0$ are uniformly distributed between $0^{\circ}$ to $180^{\circ}$.
Here $N=8100$ is the total sample number for each type of system, and $n_{\chi}$ is the number of the sample with $\chi_{\rm f}\in (\chi-d\chi/2,\chi+d\chi/2)$ and $d\chi=3^{\circ}$.
Due to the symmetry in $\chi_{\rm f}$ as seen in Figure~\ref{fig:chif}, we limit the abscissa range to $0^{\circ}-90^{\circ}$ by folding the  $\chi_{\rm f}$ values larger than $90\degr$.

From left to right, the first column of Figure~\ref{fig:chif} illustrates the probability distribution function for system A. It is clearly seen that the distribution of $\chi_{\rm f}$ is clustered around relatively small angles (less than $30\degr$). With increasing $\mu_{30,0}$, the $\chi_{\rm f}$'s distribution becomes more concentrated and the peak angle becomes smaller. This might partly explain why only a small amount of NS LMXBs show pulsations. The distribution functions for BMSPs evolved from systems B and C are presented in the second and third columns, respectively. They tend to possess large $\chi_{\rm f}$, roughly peaked around $90\degr$ when $\mu_{30,0}\leq 10$. When $\mu_{30,0}$ becomes larger, the $\chi_{\rm f}$'s distribution becomes flatter because of shorter accretion time. When $\mu_{30,0}=100$, another peak appears at $\sim 20-30\degr$. Overall, relatively large $\chi_{\rm f}$ in BMSPs are expected if $\mu_{30,0}\le 10$, which actually favors detection of pulsations from these systems. \citet{2014ApJS..213....6J} simulated the observed light curves for more than 40 MSPs detected with \textit{Fermi/LAT} and concluded that the best-fit magnetic inclination angles are almost evenly distributed between $10^{\circ}$ and $90^{\circ}$. More recently, \citet{2021A&A...647A.101B} selected several radio-loud millisecond gamma-ray pulsars showing double peaks in their gamma-ray profiles with the spin period $P_{\rm s}$ in the range of $2-6$ ms. Their best fits suggested the magnetic inclinations $\chi$ are larger than approximately $45^{\circ}$. The difference in the predicted $\chi$ distributions of UCXBs and BMSPs could be useful in testing the evolutionary models and constraining the initial parameters at the NS's birth.

We use systems D and E to simulate the formation of pulsars in ULXs. The fourth and fifth columns show similar distributions of $\chi$ when $\mu_{30,0}\leq 10$, independent of the initial magnetic moment: relatively large $\chi$s are expected for the pulsars in ULXs. When $\mu_{30,0}=100$, the $\chi_{\rm f}$'s distribution becomes flatter in system D and peaks around $30\degr$ in system E.

We emphasize that the above results are based on the assumption that the initial spin inclination angles $\alpha$ and the magnetic inclination angles $\chi$ are evenly distributed between $0\degr$ and $180\degr$. The realistic distributions must be more complicated and beyond the scope of this paper, but the general feature may not change significantly. Future observations and simulations of different types of pulsars may present more stringent constraints on their original distributions.

\section{Conclusions}\label{s:conclusion}
In this paper, we investigate the long-term magnetic inclination angle evolution of accreting NSs in I/LMXBs by combining the BA21 model with detailed binary evolution calculation. We consider five representative binary systems to reflect the formation of UCXBs, BMSPs, and ULXs. We find that the evolution of $\chi$ generally experiences at least part of the three stages: (1) during the first $10^4-10^5\,\rm yr$ mass transfer, $\chi$ does not change much due to the relatively low mass transfer rate $\dot{M}$; (2) with the growth of $\dot{M}$ and the accretion torque, $\chi$ increases and settles at its maximum after the spin inclination angle $\alpha$ evolves to $0\degr$; (3) the pulsar loss torque drives $\chi$ to decrease. We also show that stronger initial magnetic field causes the  magnetic axis to be more aligned with the spin axis for systems A, B, and C with stable mass transfer, but has little effect in systems D and E with delayed unstable mass transfer.

Moreover, considering disk instability can advance the evolution of $\chi$, but does not significantly change the final outcome except for system A. However, the enhancement of the pulsar loss torque caused by field line opening can strongly influence the evolution of $\chi$ if it really works.

Our results suggest possible distributions of the magnetic inclination angles in specific types of binary systems including UCXBs, BMSPs and ULXs. If the initial magnetic moments of NSs are moderate, BMSPs likely have relatively large magnetic inclination angles; if the NSs are initially magnetars, we expect more systems with small $\chi$ to be observed in BMSPs. Moreover, relatively small and large magnetic inclination angles are anticipated for UCXBs and ULXs, respectively.

Besides the uncertainties in both theory and observation related to the $\chi$ distribution,
the main issue in this work is that the objects are limited to  NSs in I/LMXBs, which are just a small portion of the NS population. Including other populations such as isolated NSs and those embedded in HMXBs will definitely provide more comprehensive understanding of the magnetic inclination evolution, and worth to be explored further.

\section*{acknowledgments}
This work was supported by the National Key Research and Development Program of China (2021YFA0718500), the Natural Science Foundation of China under grant No. 12041301, 12121003, and Project U1838201 supported by NSFC and CAS.

\section*{Data Availability}

The \mesa{} code, the input files necessary to reproduce our simulations, and the associated data products are available at
\dataset[zenodo.7123729]{https://doi.org/10.5281/zenodo.7123729}. The other data and codes underlying this article will be shared on reasonable request to the authors.



\begin{thebibliography}{}

\bibitem[Annala \& Poutanen(2010)]{2010A&A...520A..76A} Annala, M. \& Poutanen, J.\ 2010, \aap, 520, A76. doi:10.1051/0004-6361/200912773

\bibitem[Benli et al.(2021)]{2021A&A...647A.101B} Benli, O., P{\'e}tri, J., \& Mitra, D.\ 2021, \aap, 647, A101. doi:10.1051/0004-6361/202039853

\bibitem[Bessolaz et al.(2008)]{2008A&A...478..155B} Bessolaz, N., Zanni, C., Ferreira, J., et al.\ 2008, \aap, 478, 155. doi:10.1051/0004-6361:20078328

\bibitem[Bhattacharya \& van den Heuvel(1991)]{1991PhR...203....1B} Bhattacharya, D. \& van den Heuvel, E.~P.~J.\ 1991, \physrep, 203, 1. doi:10.1016/0370-1573(91)90064-S

\bibitem[Bhattacharyya \& Chakrabarty(2017)]{2017ApJ...835....4B} Bhattacharyya, S. \& Chakrabarty, D.\ 2017, \apj, 835, 4. doi:10.3847/1538-4357/835/1/4

\bibitem[Biryukov \& Abolmasov(2021)]{2021MNRAS.505.1775B} Biryukov, A. \& Abolmasov, P.\ 2021, \mnras, 505, 1775. doi:10.1093/mnras/stab1378

\bibitem[Bulik et al.(2003)]{2003A&A...404.1023B} Bulik, T., Gondek-Rosi{\'n}ska, D., Santangelo, A., et al.\ 2003, \aap, 404, 1023. doi:10.1051/0004-6361:20030555

\bibitem[Ghosh \& Lamb(1979)]{1979ApJ...234..296G} Ghosh, P. \& Lamb, F.~K.\ 1979, \apj, 234, 296. doi:10.1086/157498

\bibitem[Johnson et al.(2014)]{2014ApJS..213....6J} Johnson, T.~J., Venter, C., Harding, A.~K., et al.\ 2014, \apjs, 213, 6. doi:10.1088/0067-0049/213/1/6

\bibitem[Kulkarni \& Romanova(2013)]{2013MNRAS.433.3048K} Kulkarni, A.~K. \& Romanova, M.~M.\ 2013, \mnras, 433, 3048. doi:10.1093/mnras/stt945

\bibitem[Lasota(2001)]{2001NewAR..45..449L} Lasota, J.-P.\ 2001, \nar, 45, 449. doi:10.1016/S1387-6473(01)00112-9

\bibitem[Leahy(1990)]{1990MNRAS.242..188L} Leahy, D.~A.\ 1990, \mnras, 242, 188. doi:10.1093/mnras/242.2.188

\bibitem[Liu \& Li(2019)]{2019RAA....19...44L} Liu, B.-S. \& Li, X.-D.\ 2019, Research in Astronomy and Astrophysics, 19, 044. doi:10.1088/1674-4527/19/3/44

\bibitem[Long et al.(2005)]{2005ApJ...634.1214L} Long, M., Romanova, M.~M., \& Lovelace, R.~V.~E.\ 2005, \apj, 634, 1214. doi:10.1086/497000

\bibitem[Lyne \& Manchester(1988)]{1988MNRAS.234..477L} Lyne, A.~G. \& Manchester, R.~N.\ 1988, \mnras, 234, 477. doi:10.1093/mnras/234.3.477

\bibitem[Parfrey et al.(2016)]{2016ApJ...822...33P} Parfrey, K., Spitkovsky, A., \& Beloborodov, A.~M.\ 2016, \apj, 822, 33. doi:10.3847/0004-637X/822/1/33

\bibitem[Paxton et al.(2011)]{2011ApJS..192....3P} Paxton, B., Bildsten, L., Dotter, A., et al.\ 2011, \apjs, 192, 3. doi:10.1088/0067-0049/192/1/3

\bibitem[Paxton et al.(2013)]{2013ApJS..208....4P} Paxton, B., Cantiello, M., Arras, P., et al.\ 2013, \apjs, 208, 4. doi:10.1088/0067-0049/208/1/4

\bibitem[Paxton et al.(2015)]{2015ApJS..220...15P} Paxton, B., Marchant, P., Schwab, J., et al.\ 2015, \apjs, 220, 15. doi:10.1088/0067-0049/220/1/15

\bibitem[Paxton et al.(2018)]{2018ApJS..234...34P} Paxton, B., Schwab, J., Bauer, E.~B., et al.\ 2018, \apjs, 234, 34. doi:10.3847/1538-4365/aaa5a8

\bibitem[Paxton et al.(2019)]{2019ApJS..243...10P} Paxton, B., Smolec, R., Schwab, J., et al.\ 2019, \apjs, 243, 10. doi:10.3847/1538-4365/ab2241

\bibitem[Pylyser \& Savonije(1988)]{1988A&A...191...57P} Pylyser, E. \& Savonije, G.~J.\ 1988, \aap, 191, 57

\bibitem[Pylyser \& Savonije(1989)]{1989A&A...208...52P} Pylyser, E.~H.~P. \& Savonije, G.~J.\ 1989, \aap, 208, 52

\bibitem[Rankin(1990)]{1990ApJ...352..247R} Rankin, J.~M.\ 1990, \apj, 352, 247. doi:10.1086/168530

\bibitem[Ritter(1988)]{1988A&A...202...93R} Ritter, H.\ 1988, \aap, 202, 93

\bibitem[Romanova et al.(2021)]{2021MNRAS.506..372R} Romanova, M.~M., Koldoba, A.~V., Ustyugova, G.~V., et al.\ 2021, \mnras, 506, 372. doi:10.1093/mnras/stab1724

\bibitem[Romanova \& Owocki(2015)]{2015SSRv..191..339R} Romanova, M.~M. \& Owocki, S.~P.\ 2015, \ssr, 191, 339. doi:10.1007/s11214-015-0200-9

\bibitem[Shibazaki et al.(1989)]{s89} Shibazaki, N., Murakami, T., Shaham, J., \& Nomoto, K. 1989, \nat, 342, 656

\bibitem[Venter et al.(2009)]{2009ApJ...707..800V} Venter, C., Harding, A.~K., \& Guillemot, L.\ 2009, \apj, 707, 800. doi:10.1088/0004-637X/707/1/800

\bibitem[Wang \& Welter(1981)]{1981A&A...102...97W} Wang, Y.-M. \& Welter, G.~L.\ 1981, \aap, 102, 97

\bibitem[Wang(1987)]{1987A&A...183..257W} Wang, Y.-M.\ 1987, \aap, 183, 257

\bibitem[Young et al.(2010)]{2010MNRAS.402.1317Y} Young, M.~D.~T., Chan, L.~S., Burman, R.~R., et al.\ 2010, \mnras, 402, 1317. doi:10.1111/j.1365-2966.2009.15972.x

\bibitem[Zhang \& Kojima(2006)]{zk06} Zhang, C.-M. \& Kojima, Y.\ 2006, \mnras, 366, 137

\end{thebibliography}

\end{document}